\documentclass[a4,12pt,epsf]{article}
 \usepackage{graphicx}
\newcommand{\be}{\begin{eqnarray}}
\newcommand{\ee}{\end{eqnarray}}

\begin{document}

\def\descriptionlabel#1{\bf #1\hfill}
\def\description{\list{}{%
\labelwidth=\leftmargin
\advance \labelwidth by -\labelsep
\let \makelabel=\descriptionlabel}}

\begin{center}{\Large \bf
Summer Insitute 2007
\vspace{6pt}\\
Model Bulding for the Neutrino Sector and Cold Dark Matter\footnote{Talk at SI2007, Fuji-Yoshida, Japan (August2007). Dr. Y. Kajiyama and Prof. J. Kubo are my collaborator. The title is changed  a little.}}
\end{center}
 
\begin{center}
Hiroshi Okada
\vspace{6pt}\\
 
{\it Centre for Theoretical Physics at the British University in Egypt and 
Kanazawa University\\
HOkada@bue.edu.eg
}
\end{center}
\begin{abstract}
 It is now clear that the masses of the neutrino sector are much lighter than those of the other three sectors.There are many attempts to explain the neutrino masses radiatively by means of inert Higgses, which don't have vacuum expectation values. Then one can discuss cold dark matter candidates, because of no needing so heavy particles and having a $Z_2$ parity symmetry corresponding to the R-parity symmetry of the MSSM.  
   The most famous work would be the Zee model\cite{zee}.  
  
 Recently a new type model\cite{ma1}  along this line of thought was proposed by E. Ma.
We introduce a flavor symmetry based on a dihedral group $D_6$ to constrain the Yukawa sector. For the neutrino sector, we find that the maximal mixing of atmospheric neutrinos is realized, it can also be shown that only an inverted mass spectrum, the value of $|V_{MNS_{13}}|$ is 0.0034 and so on. For the fermionic CDM candidates, we find that the mass of the CDM and the inert Higgs should be larger than about $230$ and $300$ GeV, respectively.
If we restrict ourselves to a perturbative regime, they should be lighter than about $750$ GeV\cite{oka1}. 
\end{abstract}
 
\section{Model building}

 Fermionic and bosonic fields are assigned as Table1 and Table2 respectively. 

\begin{table}[thb]
\begin{center}
\begin{tabular}{|c|cccccc|} \hline
 & $L_S$ & $n_S$ & $e^c_S $&$L_I$&$n_I$&$e^c_I$ 
  \\ \hline
 $SU(2)_L\times U(1)_Y$ 
 & $({\bf 2}, -1/2)$  &  $({\bf 1}, 0)$  &  $({\bf 1}, 1)$
 & $({\bf 2}, -1/2)$&  $({\bf 1}, 0)$
 &  $({\bf 1}, 1)$
  \\ \hline
 $D_6$ & ${\bf 1}$  &  ${\bf 1}'''$  &  ${\bf 1}$
 & ${\bf 2}'$&  ${\bf 2}'$&  ${\bf 2}'$
 \\ \hline
 $\hat{Z}_2$
 & $+$ &$+$  & $-$  & $+$ 
 &$+$  & $-$
  \\ \hline
   $Z_2$
 & $+$ &$-$  & $+$  & $+$ &$-$  & $+$ 
   \\ \hline
\end{tabular}
\caption{The $D_6 \times \hat{Z}_2\times Z_2$ 
assignment for the leptons.  The subscript $S$ indicates
a $D_6$ singlet, and the subscript $I$ running from $1$ to $2$
stands for a $D_6$ doublet. $L$'s denote the $SU(2)_L$-doublet leptons,
while $e^c$ and $n$ are the $SU(2)_L$-singlet leptons.
}
\end{center}
\end{table}
\begin{table}[thb]
\begin{center}
\begin{tabular}{|c|cccc|} \hline
 & $\phi_S$ &$\phi_I$ & $\eta_S$&$\eta_I $
  \\ \hline
   $SU(2)_L\times U(1)_Y$ 
 & $({\bf 2}, -1/2)$  &  $({\bf 2}, -1/2)$   &  $({\bf 2}, -1/2)$ 
 & $({\bf 2}, -1/2)$ 
  \\ \hline
 $D_6$ & ${\bf 1}$ &${\bf 2}'$  &  ${\bf 1}'''$  & ${\bf 2}'$
 \\ \hline
 $\hat{Z}_2$ &$+$ 
 & $-$ &$+$  & $+$ 
  \\ \hline
   $Z_2$
 & $+$ &$+$  & $-$  & $-$ 
   \\ \hline
\end{tabular}
\caption{The $D_6 \times \hat{Z}_2\times Z_2$ 
assignment  for the $SU(2)_L$  Higgs doublets.
}
\end{center}
\end{table}
Under  $Z_2$  (which plays the role of $R$ parity
in the MSSM), only the right-handed neutrinos $n_S, n_I$ and 
the extra Higgs $\eta_S, \eta_I$ are odd.
The quarks are assumed to belong to ${\bf 1}$ of $D_6$ 
with $(+,+)$ of  $\hat{Z}_2\times Z_2$ so that the 
quark sector is basically the same as the SM, where 
the $D_6$ singlet Higgs $\phi_S$ with
$(+,+)$ of  $\hat{Z}_2\times Z_2$ plays the role of the SM Higgs
in this sector. No other Higgs can couple to the quark sector at the tree-level.
In this way we can avoid tree-level FCNCs in the quark sector.
So, $\hat{Z}_2$ is introduced to forbid  tree-level couplings of 
the $D_6$ singlet Higgs $\phi_S$ with the leptons
and simultaneously to forbid  tree-level couplings of $\phi_I, \eta_I$ and $\eta_S$
with the quarks.

%
\section{Lepton masses and mixing}

The most general renormalizable $D_6 \times \hat{Z}_2 \times 
Z_2$ invariant 
Yukawa interactions in the leptonic sector 
can be gained. By the Higgs mechanism, the charged lepton and the neutrino masses are generated from the $S_{2}$
invariant VEVs \cite{oka2}, and the mass matrix becomes
\begin{eqnarray}
{\bf M}_{e} = \left( \begin{array}{ccc}
-m_{2} & m_{2} & m_{5} 
\\  m_{2} & m_{2} &m_{5}
  \\ m_{4} & m_{4}&  0
\end{array}\right)
\label{mlepton}
,
{\bf M}_{\nu} = \left( \begin{array}{ccc}
2 (\rho_{2})^2 & 0 & 
0
\\ 0 & 2 (\rho_{2})^2 & 2 \rho_2 \rho_{4}
  \\ 0 & 2 \rho_2 \rho_{4}  &  
2 (\rho_{4})^2 +
(\rho_3)^2\exp i 2 \varphi_{3}
\end{array}\right),
\label{m-nu}
\end{eqnarray}
All the mass parameters
appearing in (\ref{mlepton}) can be assumed to be real.

Now we can lead some predictions for the lepton secotor.
\begin{itemize}
\item First, since the mixing of atmospheric neutrinos must be maximal form from the experiments, and only an inverted mass spectrum can be allowed.   
\item Second,\[|U_{e3}|\sim0.0034<<0.2\]
\item Third,
 \[ m_{\nu2,min.}\sim f(\tan\theta_{sol.},\Delta m^2_{32},\Delta m^2_{12},\phi=0)=0.038\sim0.067 eV,\]
\[ m_{ee,min.}\sim f(\tan\theta_{sol.},\Delta m^2_{32},\Delta m^2_{12},\phi=0)=0.034\sim0.069 eV,\]
where each of $U_{e3},m_{\nu2,min.},m_{ee,min.},\theta_{sol.},\Delta m_{32},\Delta m_{12},$
and $\phi$, which can be observed values except for  $\phi$, represents the (1-3) entry of the $Maki-Nakawgawa-Sakata$ matrix, the minimal second neutrino mass, the minimal effective majorana mass, the solar mixing angle, the atmospheric mass difference, the solar mass difference and an arbitary phase respectively.

\section{Cold Dark Matter } 
  I will move on to the discussion of the CDM .
  Where I will suppose the CDM , which is fermionic. 
Based on our model, we can consider $\mu$ $\rightarrow $ e, $\gamma$ diagram mediated only by the charged extra Higgs eta exchange. As a result of the calculation,
I find that it is more natural that ns remains as a fermionic CDM candidate.
Otherwise I have to impose a fine tuning for $n_I$ mass to sufficiently suppress the $\mu$ $\rightarrow $ e, $\gamma$ process. Furthermore I found that almostly charged extra Higgs $\eta_S$ couples to $e_L$ and $n_S$ owing to our original matrix. Therefore  there would be a clean signal if the charged extra Higgs $\eta_S$ was produced at LHC !

In the last, we would like to investigate whether or not $n_S$ can be a good CDM candidate from the cosmology. 
We found that the $n_S$ is annihilated mostly into an $e^+-e^-$ pair and a $\overline{\nu}_{\tau}-\nu_{\tau}$ pair in this model.
Reffering the following papers \cite{griest1}\cite{griest2}, we can compute the relativistic cross section.

 
 In fig. \ref{ms-ms} we present
the allowed region in the $m_S-M_S$ plane, in which
$\Omega_d h^2=0.12$
and $B(\mu\to e \gamma) < 1.2\times 10^{-11}$
are satisfied, where we assume $|h_3| <1.5$
If we allow larger $|h_3|$,
then the region expands  to
 larger $m_S$ and $M_S$,
 and for  $|h_3|\sim 0.8$ there is no allowed region.
 As we can also see from fig. \ref{ms-ms},
the mass of the CDM and the mass of the inert Higgs should be larger than 
about $230$ and $300$ GeV, respectively.
If we restrict ourselves to
a perturbative regime, they should be lighter than about $750$ GeV.
 
 \begin{figure}[htb]
 \begin{center}
\includegraphics*[width=0.60\textwidth]{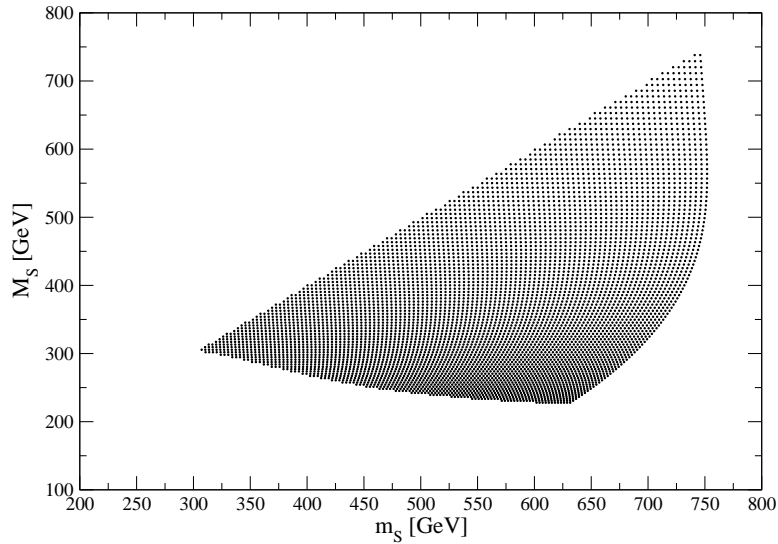}
\caption{\label{ms-ms}\footnotesize
The region in the $m_S-M_S$ plane
in which $\Omega_d h^2=0.12,
B(\mu\to e \gamma) < 1.2\times 10^{-11}$ and
$| h_3| <1.5$ are satisfied.
}
\end{center}
\end{figure}

\end{itemize}

\vspace*{12pt}
\noindent
{\bf Acknowledgement}
\vspace*{6pt}
 
\noindent
I thank organizers of SI2007, especially Prof. H. Terao . 
The Summer Institute 2007 is sponsored by
JSPS Grant-in Aid for Scientific Research (B)
No. 16340071 and also partly by the Asia
Pacific Center for Theoretical Physics, APCTP.
 

\end{document}